\documentstyle[12pt,aasms4,flushrt]{article}

\newcommand{\be}{\begin{equation}}
\newcommand{\ee}{\end{equation}}

\begin{document}

\title{POSSIBLE STELLAR METALLICITY ENHANCEMENTS 
FROM THE ACCRETION OF PLANETS}

\author{Gregory Laughlin and Fred C. Adams} 

\affil{Physics Department, University of Michigan \\
	Ann Arbor, MI 48109-1120}

\date{September 1997} 

\begin{abstract} 
A number of recently discovered extrasolar planet candidates have 
surprisingly small orbits, which may indicate that considerable
orbital migration takes place in protoplanetary systems.  A natural
consequence of orbital migration is for a series of planets to be
accreted, destroyed, and then thoroughly mixed into the convective
envelope of the central star.  We study the ramifications of planet
accretion for the final main sequence metallicity of the star.  If
maximum disk lifetimes are on the order of $\sim 10$ Myr, stars with
masses near $1.0 M_\odot$ are predicted to have virtually no
metallicity enhancement. On the other hand, early F and late A type
stars with masses $M_\ast$ $\approx$ 1.5--2.0 $M_\odot$ can experience
significant metallicity enhancements due to their considerably smaller
convection zones during the first 10 Myr of pre-main-sequence
evolution. We show that the metallicities of an aggregate of unevolved
F stars are consistent with an average star accreting $\sim 2$
Jupiter-mass planets from a protoplanetary disk having a 10 Myr
dispersal time.  
\end{abstract}

\medskip 
\noindent
{\it Subject headings:} stars: abundances, stars: stellar statistics, 
stars: planetary systems 

\section{Introduction}		\label{sec:intro}

The detection of planets orbiting nearby solar type stars (e.g., Mayor
\& Queloz 1995) was a much-anticipated discovery. The existence of
massive planets at small orbital radii ($r \leq 0.1$ AU), 
however, came as a surprise. The first such system found -- 51 Peg --
has a planetary mass $M_P$ $\sin i =$ 0.47 $M_{Jup}$ and a semi-major 
axis $r$=0.05 AU (Marcy \& Butler 1996).  Similar close companions
have been detected around $\upsilon$ And ($M_P$ $\sin i$=0.68 $M_{Jup}$,
$r$=0.057 AU), $55$ Cnc ($M_P$ $\sin i$=0.84 $M_{Jup}$, r=0.11 AU), and
$\tau$ Boo ($M_P$ $\sin i$=3.87 $M_{Jup}$, r=0.0462 AU) (see 
Butler et. al 1997).

The existence of Jupiter-mass planets very close to their primary
stars strongly suggests that orbital migration has occurred in these
systems.  The planets are posited to have formed at large radii 
(5--10 AU where most of the disk mass resides), and thereafter
transported inward as a result of tidal interactions with with the
protoplanetary disk. Successful models of orbital migration have been
constructed (e.g., Lin, Bodenheimer, \& Richardson 1996; Trilling 
et al. 1997) that demonstrate the plausibility of this hypothesis. 

As a natural consequence of the migration scenario, a series of
planetary-mass objects might be added to the star during its
pre-main-sequence development (Gonzalez 1997ab; Lin 1997; Jeffery,
Bailey, \& Chambers 1997).  Because the accreted planets are
(presumably) metal rich, the metallicity of the parent star is
enhanced.  The degree of metallicity increase depends sensitively on
the fraction of the star over which accreted planets are
distributed. A planet added to a fully convective star is folded into
the entire stellar mass and leads to a negligible metallicity
enhancement. If the parent star is largely radiative, with
only a small outer convective zone, metallicity enhancements can
be much larger. If the convective envelope subtends 2\% of the mass of
a 1 $M_{\odot}$ star, and the star accretes a Jupiter-like planet,
the metal enhancement $\Delta Z \sim$ 0.005 is both significant
and observable.

As a general trend, solar type stars begin their pre-main-sequence
evolution on the Hayashi track with a fully convective stellar
configuration. After several million years, radiative cores appear and
the size of the convective envelope steadily decreases over the next
ten million years or so. Main sequence solar mass stars have
relatively small outer convection zones which comprise only a few
percent of the stellar mass.  ZAMS stars with masses greater than
$\sim$1.3 $M_{\odot}$ are radiative up to the surface.  The expected
lifetimes for protoplanetary disks lie in the range 1--10 million
years, indicating that early F stars are likely to exhibit a
metallicity trend resulting from the accretion of planets.

This letter is organized as follows. In \S 2, we outline the basic
theoretical issues involved in determining stellar metallicity
enhancements arising from the accretion of planets. We then use these
considerations to examine a sample of nearby F stars in \S 3.  
We conclude in \S 4 by discussing our results and their implications.

\section{Metallicity Enhancement through Accretion of Planets} 
\label{sec:theory}

The metallicity of the convective envelope of a young star will evolve
as it accretes planets.  In this section, we develop a simple model of 
this process. For the sake of definiteness, we assume that the accretion 
of Jupiter-type planets is determined by an accretion function $F(t)$ 
which specifies the mass in planets accreted per unit time.  Planets,
of course, are discrete objects, so the continuous function $F(t)$
applies to an ensemble average of systems rather than particular
individual systems. The range of possibilities is adequately covered
if we use a class of accretion functions having the general form
\be 
F(t) = F_0 [ 1 - (t/t_D) ]^n \, , 
\ee
which defines a three parameter family.
The constant $F_0$ sets the overall normalization; 
the time constant $t_D$ is the total time interval over which 
planets are accreted; the index $n$ is a shape parameter that 
specifies how planet accretion falls off over time. 
The total mass in planets accreted over the entire time 
interval is given by 
\be
\Delta M = {F_0 t_D \over n + 1} \, , 
\ee
and the mean number of accreted planets is $N_P$ = $\Delta M$/$M_P$,
where $M_P$ is the mean mass of the planets. The time $t_D$ is
determined by the life time of the circumstellar disk. When the disk
mass becomes too small, orbital migration can no longer operate and
planets will no longer be accreted.  Expected disk life times are
1--10 million years, with considerable variation and uncertainty 
(see, e.g., the review of Strom, Edwards, \& Skrutskie 1993). 

Notice that rocky debris will also be accreted at some level. The
inward spiral of a gas-giant is likely to enforce the accretion of
rocky material inside its orbit (assuming the evolution is roughly
similar to that of our Solar System).  However, our numerical
experiment is limited to Jupiter-like planets.

The differential equation describing the metallicity $Z$ of the 
convective envelope of a particular star as a 
function of time can be written as
\be
{d Z \over {dt}} = (Z_P - Z) {F(t) \over M_C(M_{\ast},Z,t) } , 
\ee
where $M_C$ is the mass of the convective envelope as a function 
of time. We assume that the stellar mass $M_{\ast}$ does not change 
appreciably as planets are subsumed.  We expect that the metallicity
$Z_P$ of the planets will be comparable to that of Jupiter, and hence
$Z_P \approx 0.1$.  We also implicitly assume that the distribution 
of heavy elements in the accreted planets is the same as
the distribution of heavy elements in the Sun.

The three-dimensional function $M_C (M_{\ast},Z,t)$ must be obtained
from stellar evolutionary calculations. Using a grid of pre-main-sequence 
tracks (Forestini 1994) for model stars of mass $M_\ast$ = 1.0 -- 2.5 
$M_{\odot}$ at metallicities of $Z$ = 0.02 and 0.04, we construct the 
following analytic expression for the mass of convective envelopes  
\be 
M_C(M_{\ast},z,t)=M_{\ast}{(1-x^{2})}^{2}e^{-x^{2}} \, , 
\ee 
where we have defined 
\be 
x \equiv { t \, M_{\ast}^{3.1} \over{40(1+15(Z-0.02))}} \, . 
\ee 
This smoothly varying function reproduces the calculated values
of $M_C (M_{\ast},Z,t)$ to the required accuracy.  It provides 
accurate values for both the evolutionary time required to obtain 
a particular star's ZAMS convective envelope, and also the mass
of that envelope.

The predicted effect of accreted planets on main sequence
metallicities is illustrated in Figure 1, where several numerical
solutions to eq. (3) are plotted.  Our ``reference case'' uses
representative values $t_{D}=10$ Myr, $\Delta M = 2 M_{Jup}$, and
shape parameter $n$=2.  When planets are accreted onto a fully
radiative star, the planet's mass $M_{P}$ (taken to be 1 $M_{Jup}$ =
0.001 $M_{\odot}$) is assumed to mix with an additional mass 4$M_{P}$
of the parent star's envelope.  Figure 1 indicates that the disk
lifetime $t_D$ is the most sensitive parameter. In particular, if the
time scale for disk dispersal is $\sim1$ Myr, then no metallicity
trend from planetary accretion should be discernible in a sample of
stars in the mass range $M_\ast$ = 1.0 -- 2.5 $M_{\odot}$.

\section{A Search For a Metallicity Trend in the Nearby F Stars} 
\label{sec:obs} 

To examine the plausibility of the planetary accretion scenario, we
have used a catalog of nearby F stars (Marsakov \& Shevelev 1995;
hereafter MS) to search for trends in metallicity. The MS catalog is
based on the compilation of Hauck \& Mermilliod (1985), and contains
5489 F stars within 80 pc of the Sun.  The MS data set lists
coordinates and kinematical properties for each star, along with
metallicities (as [Fe/H]), effective temperatures, absolute
magnitudes, ages
\footnote{The ages are based on pre-HIPPARCOS parallaxes.}, 
and other data. MS determine [Fe/H] from $uvby$
photometric measurements using the calibration of Carlberg et. al
(1985).  We have found that this calibration introduces a slight
(positive) correlation between metallicity $Z$ and temperature
$T_{\rm eff}$.  We use an alternate calibration (Schuster \& Nissen 1989),
which does not exhibit such a correlation, as determined by comparing
the [Fe/H] values with measurements obtained via high resolution 
spectroscopy (Rocha-Pinto \& Maciel 1996, Wyse \& Gilmore 1995).  In
re-computing the metallicities for each star in the MS catalog, we use
$uvby$ photometry taken directly from the Hauck \& Mermilliod (1985)
compilation. The uncertainty in this method of determining 
metallicity for each star is 0.16 dex (Schuster \& Nissen 1989).

MS state that their data set is ``practically full'' in the 80 pc
limiting volume for stars having corrected colors $b-y<0.32$ (2017
stars), and 60\% full over the entire F2--G2 spectral interval (5800 K
$< T_{\rm eff} <$ 7000 K).  Because the locus of stars having $b-y=0.32$
does not correspond to a perfectly vertical line in a $Z$ vs $T_{\rm eff}$
diagram, we seek to avoid complications introduced by incompleteness
by accepting only catalog stars lying within a distance $d$ = 40 pc.
We also reject from consideration all stars belonging to unresolved
binaries.

Accreted planets generate a trend in the metallicity of stars as a
function of mass, rather than effective temperature. We must thus
determine the stellar mass as a function of metallicity and
temperature, i.e., $M_\ast (Z, T_{\rm eff})$. For a given effective
temperature and metallicity, we make bilinear interpolations from the ZAMS 
models (Forestini 1994) to obtain the stellar mass.  The MS data set is
binned according to effective temperature (all the F stars), rather
than mass.  The lower end of the $T_{\rm eff}$ range (5800--7000 K)
probes down to metal poor solar mass stars, whereas the high end of
the $T_{\rm eff}$ range includes metal rich stars with masses up to
$\sim 1.8$ $M_{\odot}$. However, only a small range in mass (1.42 --
1.52 $M_\odot$) is represented over the entire range of metallicities
in the sample.

In order to search the MS stars for a specific metallicity enrichment
trend, as predicted by eq. (3), we perform Monte-Carlo
simulations which add individual planets to an ensemble of stars in
accordance with particular choices for the accretion function. The
metallicity trend in the simulated stellar distribution can then be
compared with the trend present in the MS catalog stars.

We selected 1000 stars in the mass range 1.0 $M_{\odot}$ $< M_\ast <$
2.5 $M_{\odot}$ by sampling the initial mass function (Salpeter 1955).
These stars were then assigned metallicities in accordance with the
metallicity distribution for G dwarfs in the Solar neighborhood having
[Fe/H] $\ge -0.4$ (Rocha-Pinto \& Maciel 1996).  The metallicities in
their observed distribution were obtained from $uvby$ photometry and
the Schuster \& Nissen (1989) calibration, just as we have done.
Because G star progenitors are largely convective during the 
$\sim 10$ Myr disk dispersal time, they should not exhibit an 
overall metallicity trend from the accretion of Jupiter-type planets.
Indeed, the analysis of Rocha-Pinto \& Maciel (1996) reveals no
significant trend.

The 1000 stars were allowed to accrete planets in accordance with
the distribution given by eq. (1). The masses of individual
accreted planets were allowed to vary about the mean value of 1
$M_{Jup}$ by sampling from a gaussian distribution with variance
$M_{Jup}/3$.  The number of planets ingested by each star was sampled
from a gaussian distribution with variance $1/3$ centered on $N_P =
2$ (rounding the sample to the nearest integer). The times at which
planets were accreted were sampled directly from eq. (1), with
$n$ = 2, and $t_{D}$ = 10 Myr. The accreted planets (with $Z_P$
= 0.1) were assumed to mix thoroughly with the convective stellar
envelope, enhancing its metallicity. In the first simulation, planets
entering fully radiative stars were assumed to mix with 4 $M_{Jup}$
(0.004 $M_{\odot}$) of the sub-photospheric stellar material 
(this value is consistent with numerical estimates of G. Bryden, 
private communication 1997). 

The distribution of metallicity enhanced stars is shown in Figure 2.
After planetary accretion takes place, only 291 stars fall in the
domain shown, i.e., 6000 K $< T_{\rm eff} <$ 7000 K and 0.01 $< Z <$
0.06.  The arrows connect the ($Z$,$T_{\rm eff}$) value that each star
would have with no planet accretion with the ($Z$,$T_{\rm eff}$) value
obtained after including planet accretion.  The figure also shows the
zero-age F stars in the MS catalog that have distances $d<40$ pc and
fall within the plot limits (305 stars).  We have used only unevolved
stars (MS catalog stars whose ages are consistent with a ZAMS
identification) in order to minimize the complications brought on by
both main sequence stellar evolution and the overall metallicity
evolution of the Milky Way.

A linear fit to the original metallicities of the 291 simulation stars
shows a slight positive slope of $0.6 \pm 0.8 \times 10^{-6} {\rm
K}^{-1}$.  A linear fit to the distribution of enhanced metallicities
shows a larger slope of $2.5 \times 10^{-6} {\rm K}^{-1}$, which is
consistent and comparable with the slope of $1.8 \pm 1.7 \times
10^{-6} {\rm K}^{-1}$ obtained from the 305 MS catalog stars.

In Figure 3, we show the results of a simulation which contains the
same parameters as the simulation described above, except that we have
changed the minimum mixing mass to 2 $M_{Jup}$.  Notice that this 
simulation predicts too many high metallicity stars compared to the 
MS data set. We can thus conclude that accreted planets must mix 
with more than 2 $M_{Jup}$ of stellar material, or that less than 
2 planets per star can be accreted. 

\section{Discussion} \label{sec:sum} 

We have studied the possible enhancement of stellar metallicity due to
the accretion of planets.  In particular, we have developed a simple
theoretical model to incorporate metal enhancements due to accreted
planets.  We have then searched for trends in metallicity within a
well defined sample of nearby F stars. Our results are summarized
below.

[1] The effects of planet accretion on stellar metallicity are
strongly dependent upon the total stellar mass.  Stars with masses
comparable to the Sun have large convective envelopes for nearly the
entire time interval over which planets are expected to be accreted. 
These stars will show essentially no metal enhancements. Higher mass
stars, with $M_\ast \sim 1.5 M_\odot$, have much smaller convective
envelopes during their pre-main-sequence phases and can suffer
relatively larger metal enhancements. We have found that the disk
lifetime is the most important parameter in determining the
metallicity trend. Disk lifetimes $t_{D}>$ 10 Myr can lead to
significant metallicity increases for F stars which accrete planets.

[2] The MS catalog of nearby F stars shows only marginal evidence for
a trend of increasing metallicity with increasing stellar mass. Such a
trend is consistent with the metallicity increases expected from the
accretion of about 2 Jupiter-mass planets per star over a disk
dispersal lifetime of 10 Myr.  This work can only point out the 
{\it possible} existence of this trend.  A more refined analysis 
using a larger sample of stars and/or more accurate metallicity 
determinations is required to definitely establish or rule out 
its existence. 

Indeed, a great deal of follow-up work can be done to elucidate these
concepts further.  Because the predicted metallicity trends resulting
from planetary accretion are fairly subtle, a larger and more complete
sample of stars, with an increased mass range, is needed to obtain the
statistics necessary for a firm conclusion.  The trend is predicted to
be most significant for stars in the mass range 1.5 -- 2.5 $M_\odot$.
It is also highly advantageous to obtain the stellar metallicities
from spectroscopy, rather than the cruder measure afforded by 
{\it uvby} photometry. 
\footnote{We note that the spectroscopic metallicity estimates of 
Edvardsson et al. (1993) are smaller than the photometric metallicity 
estimates for the most metal-rich stars (see also Wyse \& Gilmore 1995).} 
It is possible that an unforeseen effect is influencing the $b-y$ index 
as the spectral type changes.  However, one additional complication is 
that metal lines are more difficult to observe for A stars. 
Another possible improvement is to recalibrate the Schuster \& Nissen 
equation with recent spectroscopic metallicity estimates.  

Notice that this approach -- searching for metallicity trends in a
wide sample of nearby stars -- is complementary to directly studying
the specific systems which are thought to have planets. The stars in
these newly discovered planetary systems seem to be systematically
high in metallicity (e.g., Gonzalez 1997ab).  For small numbers of
systems, however, it is difficult to separate the effects of
metallicity enhancement from high initial metallicity.  We note, for
example, that planetary accretion does not account for the high
metallicities of 55 Cnc and 51 Peg; these G stars should have large
convection zones throughout the planet accretion phase of evolution.
To circumvent the initial metallicity ambiguity, one can search for
metallicity trends among F and G dwarfs in open clusters and binary
systems. Wide binaries containing both an early G dwarf and an early F
dwarf will thus provide another good test of this scenario.

Additional theoretical work also remains.  A comprehensive set of
pre-main-sequence evolutionary tracks should be computed with varying
core and envelope metallicities, and an emphasis on the mass of the
outer convection zone with time.  Heavy elements settle out of stellar
photospheres if given enough time and these effects can also be 
included.  Further refinements to the orbital migration problem should
be done. Finally, the manner in which a giant planet spirals into a
stellar photosphere should be studied in more detail.
 
\acknowledgements

We are pleased to thank P. Bodenheimer for useful discussions and
a critical reading of the manuscript. We are grateful to G. Gonzalez for
providing a prompt, insightful, and useful referee's report.
This work was supported by a
NSF Young Investigator Award, NASA Grant No.~NAG~5-2869, and by funds
from the Physics Department at the University of Michigan.

\begin{center}
\large Figure Captions
\end{center}

\figcaption{
Enrichment trends for stars of mass $1.0 M_{\odot} < M_\ast < 2.5
M_{\odot}$.  The reference case ({\it thick gray line}) shows the
metallicity trend for a populations of stars with disk lifetimes
$t_D$ = 10 Myr, accretion index $n$=2, and 2 Jupiter
masses accreted. The additional thin lines show the effect of varying
each of these three parameters individually: The {\it short dashed
lines} show trends expected from $N_{Jup}$ = 1, 3.  {\it Solid lines}
show trends expected for disk lifetimes of 20, 15, 5 and 1 Myr. 
{\it Long dashed} lines show trends expected for $n$ = 1 and 3.
\label{fig1}}

\figcaption{ 
Monte-Carlo simulation of metallicity enrichment due to
accretion of planets.  Metal enriched stars (obtained as described in
the text) are shown as black dots. Arrows indicate the change in
$T_{\rm eff}$ and $Z$ resulting from metal enrichment.
The solid black line shows a 
linear fit to the enhanced metallicity population, the dashed line
shows a linear fit to the original distribution of metallicities and
effective temperatures.  Zero age F stars lying within a distance $d$
= 40 pc of the Sun are shown as gray open circles (from the MS
catalog).  The solid gray line shows a linear fit to the metallicities
of these stars. The minimum mixing mass beneath the stellar
photosphere is 4 $M_{Jup}$. The triangles indicate average metallicity
values within 200K temperature bins.
\label{fig2}}

\figcaption{
Same as Figure 2 except that the minimum mixing mass beneath the stellar
photosphere is 2 $M_{Jup}$.
\label{fig3}}

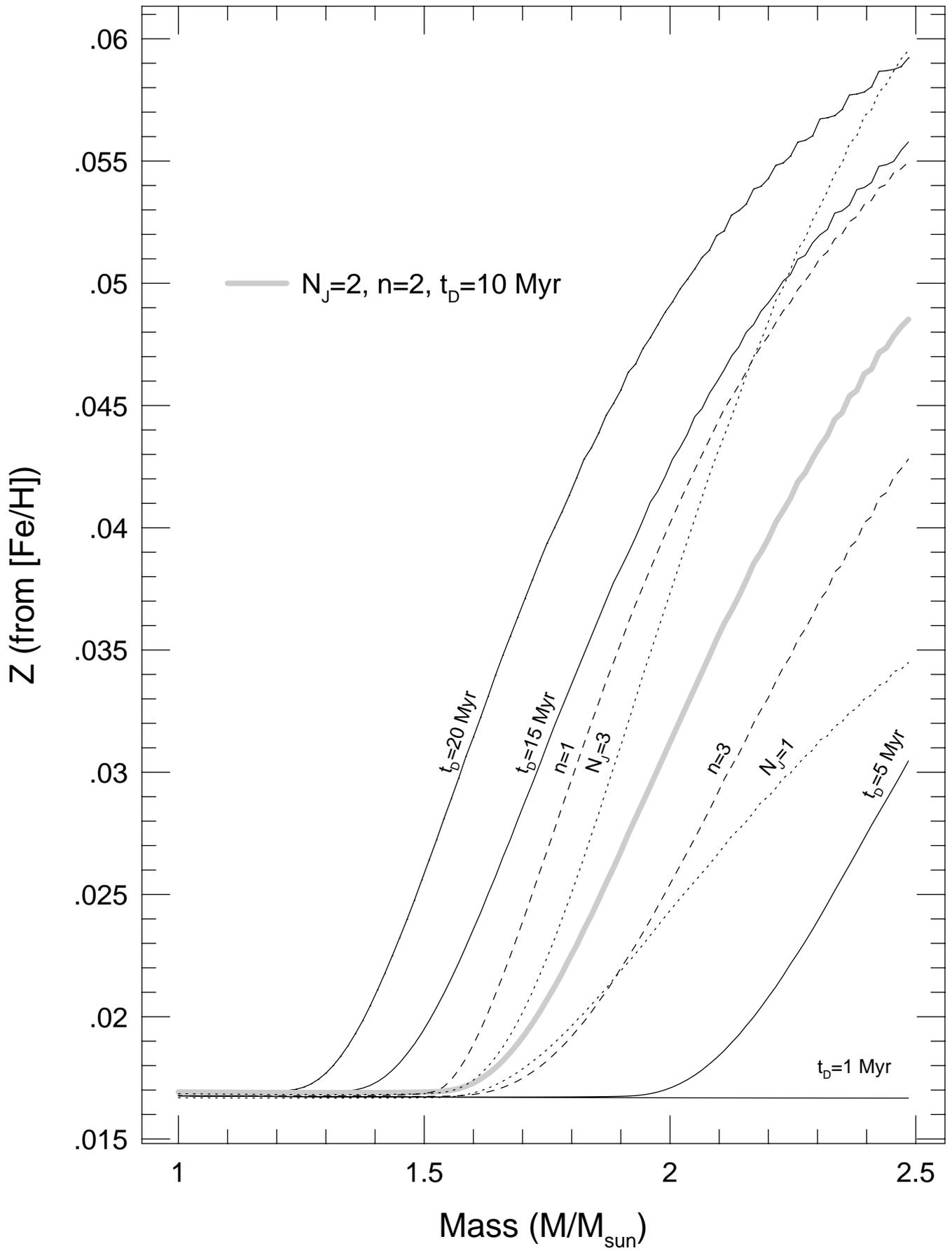
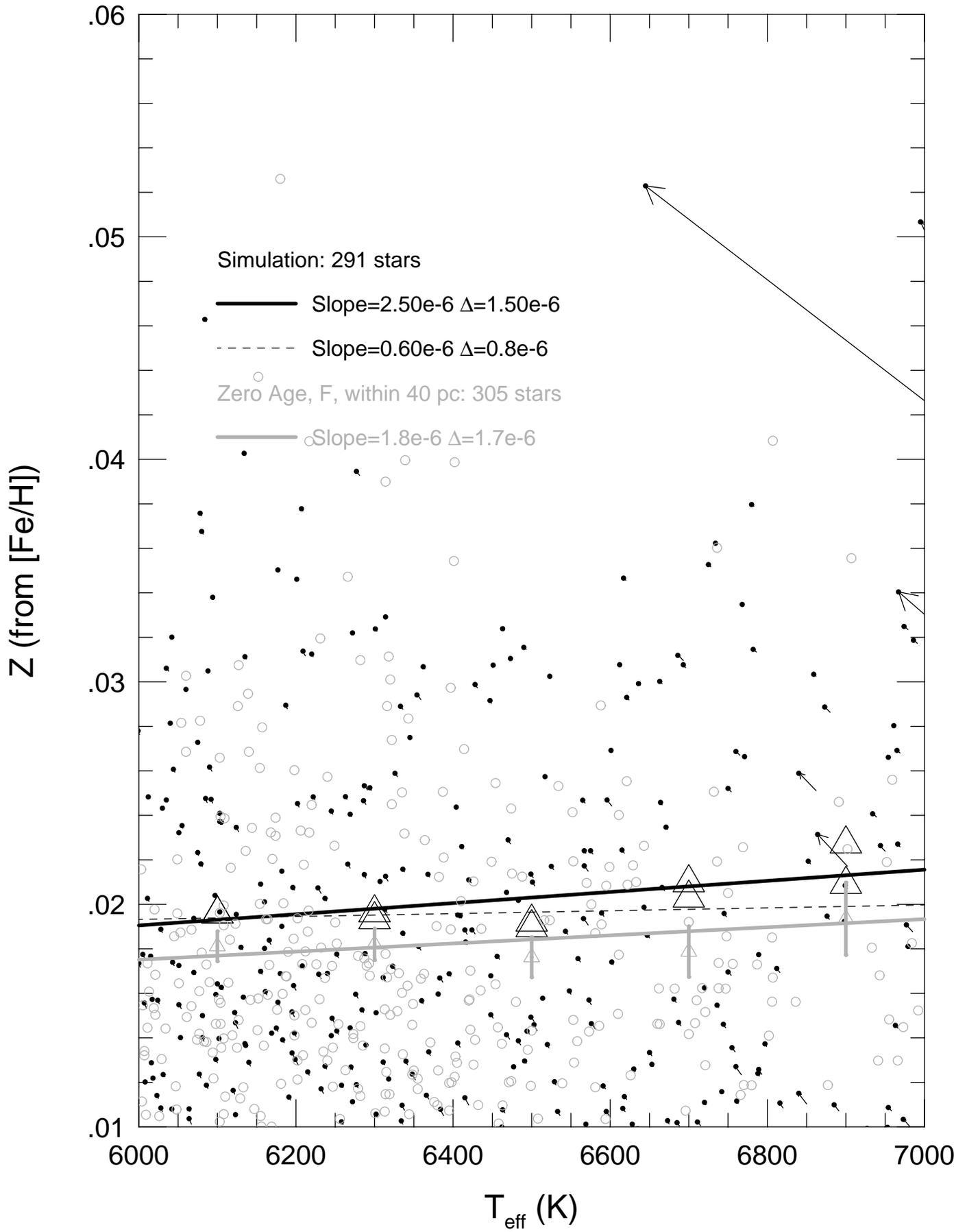
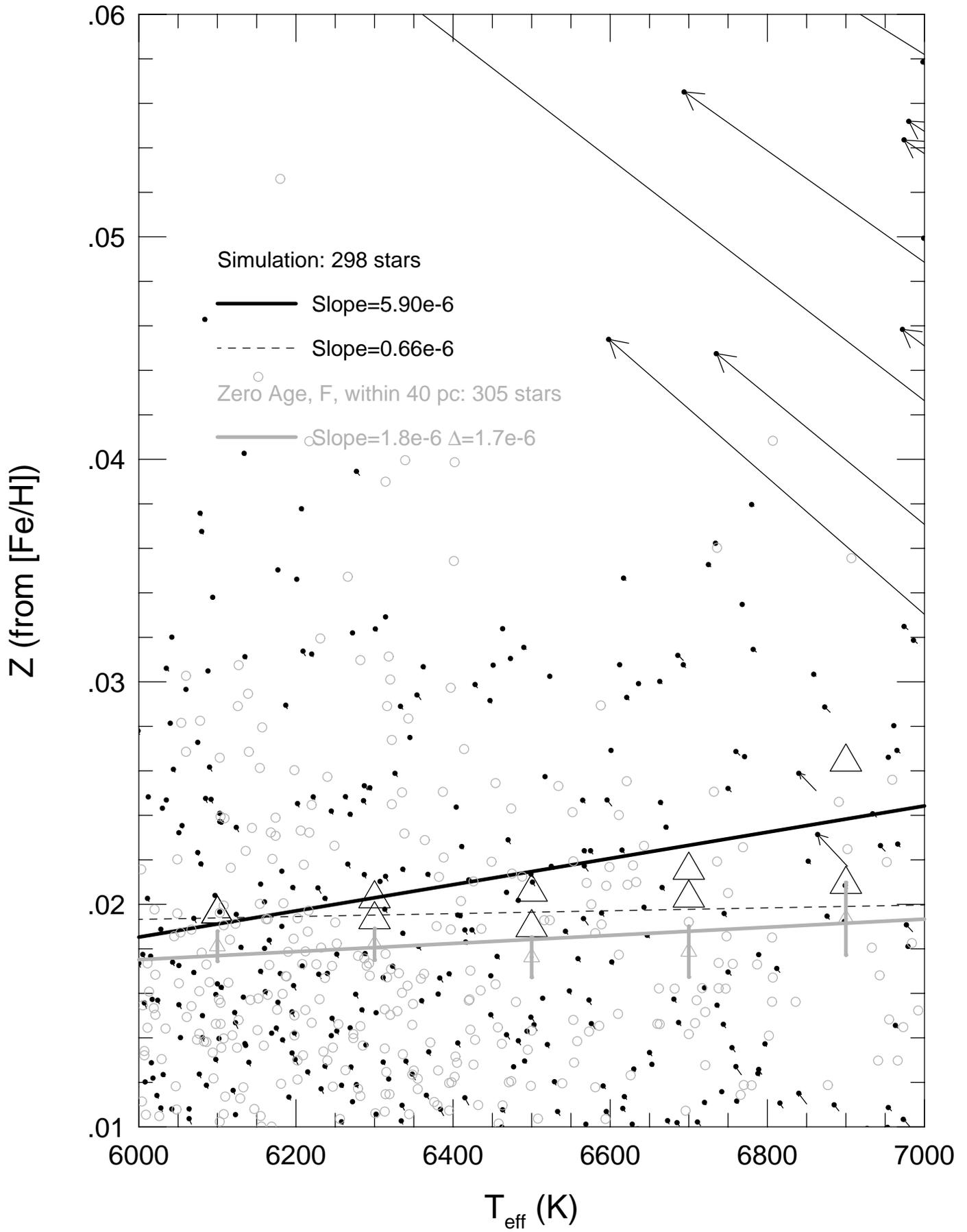
\end{document}